\documentclass[journal,12pt,onecolumn,draftclsnofoot,]{IEEEtran}
\usepackage[british]{babel}
\usepackage[ruled, vlined]{algorithm2e}
\usepackage{algpseudocode}
\usepackage{amsmath}
\usepackage{amssymb}
\usepackage{array}
\usepackage{arydshln}
\usepackage{blkarray, bigstrut}
\usepackage{bm}
\usepackage{booktabs}
\usepackage{braket}
\usepackage{cite}
\usepackage{enumitem}
\usepackage{graphicx}
\usepackage[mathscr]{euscript}
\usepackage{mathtools}
\usepackage{multirow}
\usepackage{framed} 
\usepackage[framed]{ntheorem}
\usepackage{nicefrac}
\usepackage{pgfplots}
\usepackage{qtree}
\usepackage{adjustbox}
\usepackage{rotating}
\pgfplotsset{compat = 1.13}
\usepackage{stfloats}
\usepackage{subfig}
\usepackage{url}
\usepackage{tikz}
\usepackage{tkz-berge}
\usetikzlibrary{patterns}
\usetikzlibrary{spy}
\usetikzlibrary{shapes, backgrounds,calc, snakes}
\usetikzlibrary{decorations.pathreplacing}
\usetikzlibrary{matrix}
\usepackage{fancybox}

\tikzstyle{vertex} = [circle, draw, inner sep = 0pt, minimum size = 10pt]

\newframedtheorem{frm-prop}{Property}

\definecolor{bblue}{rgb}{0.12392, 0.0490, 0.9588}
\definecolor{sskyblue}{rgb}{0.1529, 0.5882, 0.9216}
\definecolor{ggreen}{rgb}{0.5020, 0.7961, 0.3451}
\definecolor{yyellow}{rgb}{0.9765, 0.9804, 0.0784}

\definecolor{color0}{HTML}{FF0147}
\definecolor{color1}{HTML}{F400DC}
\definecolor{color2}{HTML}{BA0DFF}
\definecolor{color3}{HTML}{5700E8}
\definecolor{color4}{HTML}{0B03FF}
\definecolor{color5}{HTML}{0957F4}
\definecolor{color6}{HTML}{03B3FF}
\definecolor{color7}{HTML}{08E8DA}
\definecolor{color8}{HTML}{07FF8E}
\definecolor{color9}{HTML}{51FF0A}

\definecolor{p1}{rgb}{1, 0.0667, 0}
\definecolor{p2}{rgb}{1, 0.24, 0}
\definecolor{p3}{rgb}{1, 0.349, 0}
\definecolor{p4}{rgb}{1, 0.490, 0}
\definecolor{p5}{rgb}{1, 0.631, 0}
\definecolor{p6}{rgb}{1, 0.792, 0}
\definecolor{p7}{rgb}{1, 0.933, 0}

\begin{document}

\title{Network-Assisted Resource Allocation with Quality and Conflict Constraints for V2V Communications}
\author{Luis F. Abanto-Leon, Arie Koppelaar, Sonia Heemstra de Groot}

\maketitle

\begin{abstract}
	The 3rd Generation Partnership Project (3GPP) has recently established in Rel. 14 a network-assisted resource allocation scheme for vehicular broadcast communications. Such novel paradigm is known as vehicle--to--vehicle (V2V) \textit{mode-3} and consists in eNodeBs engaging only in the distribution of sidelink subchannels among vehicles in coverage. Thereupon, without further intervention of the former, vehicles will broadcast their respective signals directly to their counterparts. Because the allotment of subchannels takes place intermittently to reduce signaling, it must primarily be conflict-free in order not to jeopardize the reception of signals. We have identified four pivotal types of allocation requirements that must be guaranteed: one quality of service (QoS) requirement and three conflict conditions which must be precluded in order to preserve reception reliability. The underlying problem is formulated as a  maximization of the system sum-capacity with four types of constraints that must be enforced. In addition, we propose a three-stage suboptimal approach that is cast as multiple independent knapsack problems (MIKPs). We compare the two approaches through simulations and show that the latter formulation can attain acceptable performance at lesser complexity.
\end{abstract}

\begin{IEEEkeywords}
	subchannel allocation, broadcast vehicular communications, quality of service
\end{IEEEkeywords}

\IEEEpeerreviewmaketitle

\section{Introduction}
The 3rd Generation Partnership Project has recently proposed in Release 14 two new resource allocation concepts for vehicle--to--vehicle (V2V) communications, namely V2V \textit{mode-3} and V2V \textit{mode-4}. The latter one, is intended for supporting scenarios wherein network coverage is not available. Thus, vehicles will be required to sense the occupancy of subchannels and reserve a subset for their own transmission. Each vehicle reserves subchannels in a semi-persistent manner while attempting not degrade the link conditions of their counterparts \cite{b1}. On the other hand, in V2V \textit{mode-3} eNodeBs only provide support in the apportionment of subchannels but do not intervene in traffic control as occurs with mainstream cellular communications. Thus, once subchannels have been distributed, vehicles will broadcast their signals in turns over the designated resources until a new allocation is processed \cite{b2}. Because the assignment of subchannels does not take place frequently (e.g. once every few hundred milliseconds or more) to reduce signaling information, it must be $(i)$ conflict-free and $(ii)$ provide sufficient capacity to satisfy the differentiated quality of service (QoS) requirements for each vehicle. 

In this work, we provide a formulation for the described subchannel allocation problem in V2V \textit{mode-3}. The objective is to maximize the sum-capacity of the system---consisting of several vehicles distributed over a number of clusters---while enforcing the fulfillment of four types of requirements which are described in more detail in the following section. Moreover, we propose a simplified three-stage formulation of the primal problem in terms of multiple independent knapsack problems (MIKPs) \cite{b3}. In the initial stage, the clusters are hierarchically sorted based on their cardinality. In the second stage, every vehicles from each cluster is matched with time-domain subframe. In the last stage, vehicles are apportioned specific subchannels from within the assigned subframe such that the QoS requirements are fulfilled. Across all the stages, subchannels are selected such that conflicts of any type are prevented. 

The remaining of the paper is organized as follows. In Section II, we enunciate the motivation of the present work and briefly summarize our contributions. In Section III, the subchannel allocation problem for V2V \textit{mode-3} including four types of constraints is formulated. In Section IV, a simplified allocation approach based on MIKPs is described. Section V discusses simulation results and Section VI is devoted to summarizing our concluding remarks.

\section{Motivation and Contributions}
\begin{figure*}
	\begin{center}
		\begin{tikzpicture}
		\node (img) {\includegraphics[width=0.85\linewidth]{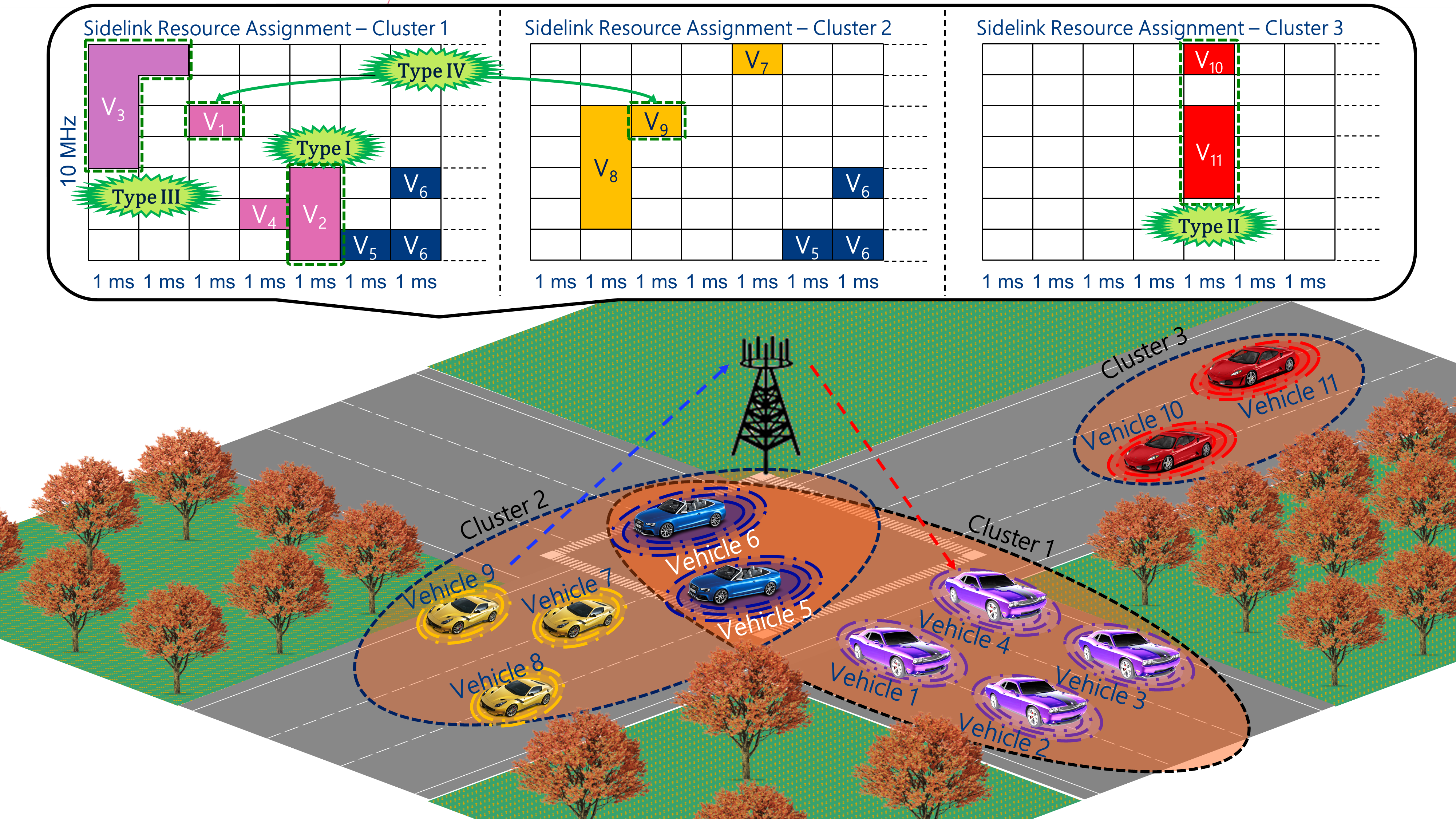}};
		\end{tikzpicture}
		\caption{Vehicular broadcast communications in V2V \textit{mode}-3}
		\label{f1}
		\vspace{-0.5cm}
	\end{center}
\end{figure*}

Fig. \ref{f1} depicts an scenario with $N = 11$ vehicles grouped into 3 clusters. It can be observed that \textit{cluster 1} consists of vehicles $\{ v_1, v_2, v_3, v_4, v_5, v_6 \}$, \textit{cluster 2} consists of vehicles $\{ v_5, v_6, v_7, v_8, v_9 \}$ whereas \textit{cluster 3} is constituted by vehicles $\{ v_{10}, v_{11} \}$.  Moreover, vehicles $\{ v_5, v_6 \}$ lie at the intersection of \textit{cluster 1} and \textit{cluster 2}. Depending on the distribution of vehicles across the clusters, some subchannel assignments might be detrimental since interference could be originated due to subchannel repurposing and thus impinging on communication reliability. We have identified four types of conditions that are compulsory for QoS-aware conflict-free allocations in V2V broadcast communications \cite{b6}.

\begin{itemize}
	\item \textit{Type I}: Each vehicle has a differentiated QoS requirement. We define QoS in terms of the channel capacity required by a vehicle to convey the intended signal. For instance, in Fig. \ref{f1}  vehicles $v_8$ and $v_6$ have been assigned four and two subchannels, respectively.
	\item \textit{Type II}: When two or more vehicles in the same cluster transmit concurrently, they cannot receive the signals of the others due to half-duplex PHY. This problem does not affect other vehicles in the cluster, only those that engaged in simultaneous transmission. An example of this case is depicted by $v_{10}$ and $v_{11}$, which have been assigned subchannels in the same subframe.
	\item \textit{Type III}: In order to support scenarios with high vehicular density and improve the utilization of radio resources, the subchannels assigned to a vehicle should preferably be selected from within the same subframe. This problem is depicted by vehicle $v_3$ whose allotted subchannels span two subframes.
	\item \textit{Type IV}: Vehicles lying at the intersection of clusters may receive concurrent signals from other vehicles that are not aware of each other, similarly to the hidden node problem experienced in IEEE 802.11p \cite{b4}. Thus, signals from different vehicles may overlap concurrently in time and frequency; thus becoming undecodable to other vehicles. For example, if vehicles $v_1$ and $v_9$ transmit in the same subchannel, vehicles $v_5$ and $v_6$ may not be able to correctly decode the received signals.
\end{itemize}

Considering the requirements above described, we propose a formulation for the subchannel allocation problem in V2V \textit{mode-3}. Furthermore, we propose a simplified method consisting of three stages. In the first stage, the clusters\footnote{The clusters are formed based on their similarity in position, speed and direction using an affinity Gaussian kernel as described in \cite{b8}.} are sorted based on their cardinality, from the highest to the lowest. Thus, clusters with a higher number of vehicles will be processed first as these are more complicated to optimize. In the second stage, we perform a matching between vehicles and subframes in a random manner without explicitly allotting specific subchannels to vehicles. This prevents Type III conflicts as the subchannels (to be selected in the third stage) will be confined to a single subframe. In addition, all the vehicles that belong to the same cluster must be placed in different subframes to prevent Type II conflicts. In the third stage, a method for attaining the differentiated QoS requirements of Type I for each vehicle is formulated as multiple independent knapsack problems (MIKPs). Since the allocation of subchannels for the whole system is performed sequentially for each cluster at a time, subchannels that may cause vehicles at the intersection to undergo Type IV conflicts can be removed from the apportionment process. Thus, all the four types of requirements can be fulfilled. We show through simulations that the proposed approach based on MIKPs can attain acceptable performance when compared to the optimal solution.

\section{Problem Formulation}
We consider that downlink and uplink spectrum resources are available for periodical signaling between vehicles and the eNodeB. In-coverage vehicles will report via uplink information on the perceived quality of subchannels. Based on such information, the eNodeB will assign subchannels to each vehicle. Moreover, via downlink the eNodeB informs the vehicles of the designated subchannels for their use. For the system, we have considered a 10 MHz channel for exclusive sidelink broadcast communications between vehicles.

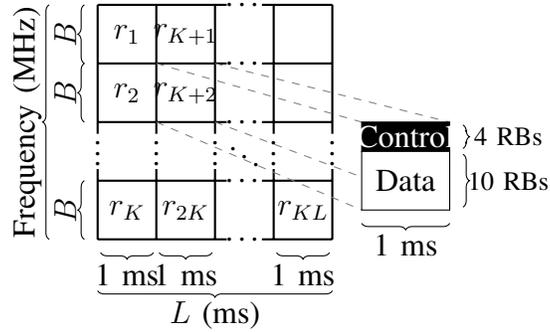
\begin{figure}[!t]
	\centering
	\begin{tikzpicture}[scale = 0.78]
	\draw[step=1cm, thick] (0,0) grid (2.2,-2.2);
	\draw[step=1cm, thick] (2.8,0) grid (4,-2.2);
	\draw[step=1cm, thick] (0,-2.8) grid (2.2,-4.001);
	\draw[step=1cm, thick] (2.8,-2.8) grid (4,-4.001);
	
	\draw [color=gray, dashed] (1,-1) -- (4.5,-2);
	\draw [color=gray, dashed] (2,-1) -- (6,-2);
	\draw [color=gray, dashed] (1,-2) -- (4.5,-3.5);
	\draw [color=gray, dashed] (2,-2) -- (6,-3.5);
	
	\draw[fill=black] (4.5,-2) rectangle (6,-2.5);
	\node at (5.25,-2.25) {\textcolor{white}{Control}};
	\draw[fill=white] (4.5,-2.5) rectangle (6,-3.5);
	\node[fill = white] at (5.25,-3) {\textcolor{black}{Data}};
	
	\draw[decoration={brace, raise=5pt},decorate] (6, -2.05) -- node[right=6pt] {} (6, -2.45);
	\node [align=left] at (7, -2.25) {\small 4 RBs};
	\draw[decoration={brace, raise=5pt},decorate] (6, -2.55) -- node[right=6pt] {} (6, -3.45);
	\node [align=left] at (7, -3) {\small 10 RBs};
	\draw[decoration={brace, raise=5pt},decorate] (6,-3.5) -- node[right=6pt] {} (4.5,-3.5);
	\node at (5.25,-4.1) {1 ms};
	
	\node at (2.5,0) {\dots};
	\node at (2.5,-1) {\dots};
	\node at (2.5,-2) {\dots};
	\node at (2.5,-3) {\dots};
	\node at (2.5,-4) {\dots};
	
	\node at (0, -2.4) {\vdots};
	\node at (1, -2.4) {\vdots};
	\node at (2, -2.4) {\vdots};
	\node at (3, -2.4) {\vdots};
	\node at (4, -2.4) {\vdots};
	
	\node at (2.5, -2.45) {$\ddots$};
	
	\draw[decoration={brace, raise=5pt},decorate] (0.95, -4.0) -- node[right=6pt] {} (0.05, -4.0);
	\draw[decoration={brace, raise=5pt},decorate] (1.95, -4.0) -- node[right=6pt] {} (1.05, -4.0);
	\draw[decoration={brace, raise=5pt},decorate] (3.95, -4.0) -- node[right=6pt] {} (3.05, -4.0);
	
	\draw[decoration={brace, raise=5pt},decorate] (0, -1) -- node[right=6pt] {} (0.0, 0);
	\draw[decoration={brace, raise=5pt},decorate] (0, -2) -- node[right=6pt] {} (0.0, -1);
	\draw[decoration={brace, raise=5pt},decorate] (0, -4) -- node[right=6pt] {} (0.0, -3);

	\draw[decoration={brace, raise=5pt},decorate] (4, -4.65) -- node[right=6pt] {} (0.0, -4.65);
	\node at (2,-5.3) {$L$ (ms)};
	
	\draw[decoration={brace, raise=5pt},decorate] (-0.6, -4) -- node[right=6pt] {} (-0.6, 0);
	\node[rotate = 90] at (-1.2,-2) {Frequency (MHz)};
	
	\node at (0.5,-4.6) {1 ms};
	\node at (1.5,-4.6) {1 ms};
	\node at (3.5,-4.6) {1 ms};
	
	\node at (0.5,-0.5) {$r_{1}$};
	\node at (0.5,-1.5) {$r_{2}$};
	\node at (0.5,-3.5) {$r_{K}$};
	
	\node at (1.5,-0.5) {$r_{K+1}$};
	\node at (1.5,-1.5) {$r_{K+2}$};
	\node at (1.5,-3.5) {$r_{2K}$};
	
	\node at (3.5,-3.5) {$r_{KL}$};
	
	\node[rotate = 90] at (-0.55,-0.5) {$B$};
	\node[rotate = 90] at (-0.55,-1.5) {$B$};
	\node[rotate = 90] at (-0.55,-3.5) {$B$};
	
	\end{tikzpicture}	
	\caption{Sidelink subchannels for V2V communications}
	\label{f2}
	\vspace{-0.25cm}
\end{figure}

The channelization of spectrum resources into subchannels to serve in sidelink communications \cite{b5} is shown in Fig. \ref{f2}. The number of subframes is denoted by $L$ and each consists of $K$ subchannels of duration 1 ms and bandwidth $B = 1.26$ MHz such that $KB \leq 10$ MHz. Thus, in each subframe, at most 7 subchannels can be supported. The reason for such granularity is that in safety applications, a single subchannel with the specified dimensions, i.e. 14 resource blocks (RBs) can sufficiently bear a CAM message. However, for other types of applications, a larger amount of subchannels might be required. In Fig. \ref{f2}, $r_k$ represents a sidelink subchannel and $\mathcal{R}_l = \{ r_{(l-1)K + 1}, \dots,  r_{lK}\}$ is the set of subchannels contained in subframe $l$, for $l = 1, 2, \dots, L$. Thus, the whole set of subchannels in an allocation window of $L$ ms is given by $\mathcal{R} = \cup_{l=1}^L \mathcal{R}_l = \{r_1, r_2, \dots, r_{KL} \} $. Also, the total number of vehicles distributed among $J$ clusters in the system is denoted by $N$. Thus, if $\mathcal{V}^{(j)}$ denotes a particular cluster $j$, then $\mathcal{V} = \cup_{j} \mathcal{V}^{(j)}= \{v_1, v_2, \dots, v_N \}$ represents all the vehicles in the system. On the other hand, $x_{ik}$ is a boolean variable that indicates with 1 whether a vehicle $v_i \in \mathcal{V}$ and subchannel $r_k \in \mathcal{R}$ are matched or with 0 otherwise. Also, the achievable capacity that vehicle $v_i$ can attain if it transmits in subchannel $r_k$ is represented by $c_{ik} = B \log_2(1 + \mathsf{SINR}_{ik})$. Similarly, $\mathsf{SINR}_{ik}$ is the signal--to--interference--plus--noise ratio (SINR) that vehicle $v_i$ perceives in subchannel $r_k$\footnote{In strict sense, $c_{ik}$ and $\mathsf{SINR}_{ik}$ depict metrics between vehicle $v_i$ and some vehicle $v_u$ with which $v_i$ experiences the weakest link quality. Thus, if the weakest link can be leveraged, other vehicles receiving signals from $v_i$ may experience superior conditions. For the purpose of simplicity, the index $u$ representing vehicle $v_u$ has been dropped.}. 

In the forthcoming subsections, the objective function for the subchannel allocation problem is introduced. Then, the four types of assignment requirements are described in detail.

\subsection{Objective Function}
The aim is to maximize the sum-capacity of the system while satisfying the four types of allocation requirements. The objective function is expressed as ${\bf c}^T {\bf x}$ where ${\bf x}$ and ${\bf c}$ are vectors containing elements $x_{ik}$ and $c_{ik}$ for all the vehicles $v_i$ and subchannels $r_k$, i.e. $\mathbf{x} = [x_{1,1}, \dots, x_{1,KL}, \dots, x_{N,1}, \dots, x_{N,KL}]^T$, $\mathbf{c} = [c_{1,1}, \dots, c_{1,KL}, \dots c_{N,1}, \dots, c_{N,KL}]^T$ 

\subsection{Type I: Per-vehicle QoS requirement}
For each vehicle $v_i$ the required capacity to transmit the intended signal is denoted by $q_i$. Such demand is expressed by $ \sum_{k = 1}^{KL} c_{ik} x_{ik} = q_i$, for $i = 1, 2, \dots, N$. Since fulfillment of the exact requested $q_i$ may not be feasible, the condition can be slightly relaxed and cast as $ q_i - \epsilon \leq \sum_{k = 1}^{KL} c_{ik} x_{ik} \leq q_i + \epsilon$. Thus, for all the $N$ vehicles in the system, the set of constraints can be expressed as
\begin{equation} \label{e1}
	{\bf q}_{N \times 1} - {\bm \epsilon} \leq ({\bf I}_{N \times N} \otimes {\bf 1}_{1 \times KL})({\bf c} \circ {\bf x}) \leq {\bf q}_{N \times 1} + {\bm \epsilon}
\end{equation}
where ${\bf q} = [q_1, q_2, \dots, q_N ]^T$ and ${\bm \epsilon} = \epsilon \cdot ({\bf 1}_{N \times 1}) $, $\exists ~\epsilon \geq 0$. The symbols $\otimes$ and $\circ$ represent the Kronecker and Hadamard product, respectively. 

\subsection{Type II: Intra-cluster subframe allocation conflicts}
When two or more intra-cluster vehicles transmit in subchannels that belong to the same subframe, they will not able to receive each other's signals due to half-duplex PHY assumption. This, however, will not affect other vehicles. If we can guarantee that no pair of intra-cluster vehicles will transmit concurrently in subchannels of the same subframe, this kind of conflict can be avoided. Let $v_y$ and $v_z$ denote two different vehicles in the same cluster, thus when the condition $\left( \sum_{k_1} x_{y{k_1}} \right) \left( \sum_{k_2} x_{z{k_2}} \right) = 0$ holds, conflicts do not occur. The indexes $k_1$ and $k_2$ represent two subchannels $r_{k_1}$ and 
$r_{k_2}$ belonging to the same subframe $\mathcal{R}_{l'}$, for $l' = 1, 2, \dots, L$. The equality is non-zero only when both vehicles transmit on at least one subchannel of $\mathcal{R}_{l'}$. More generally, for $N$ vehicles, a compact form of expressing these constraints is given by  
\begin{equation} \label{e2}
	[({\bf G}_{P \times N}^{+} \otimes {\bf I}_{L \times L}) {\bf x}_s] \circ [({\bf G}_{P \times N}^{-} \otimes {\bf I}_{L \times L}) {\bf x}_s] = {\bf 0}_{PL \times 1}
\end{equation}
where ${\bf x}_s = ({\bf I}_{NL \times NL} \otimes {\bf 1}_{1 \times K}) {\bf x}$. The total number of intra-cluster vehicle pairs in the whole system is denoted by $P$. The boolean matrices ${\bf G}^{+}$ and ${\bf G}^{-}$ have a strong relation with the topology of the scenario and the distribution of vehicles across the clusters. These matrices also collect information on the restricted allocations that lead to this kind of conflict.

\subsection{Type III: Minimal time-dispersion of subchannels}
When the subchannels apportioned to a vehicle span over several subframes, the signal duration over the air persists longer. And because signals broadcasted by vehicles are of periodic nature, this implies that less time will remain for other vehicles (in the same cluster) to transmit. Therefore, if a vehicle has high QoS requirements, subchannels should be selected from within the same subframe since this will allow to maximize the number of served vehicles. Moreover, considering the described channelization, vehicles can be assigned up to $K$ subchannels from any subframe. For any vehicle $v_i$, to guarantee that the allotted subchannels will be confined to a single subframe, the following must hold $ (\sum_{u \in \mathcal{R}_{l}} x_{iu}) (\sum_{u' \in \mathcal{R}_{l'}} x_{iu'}) = 0$, for $l \neq l'$ $\forall~l, l' = 1, 2, \dots, L$. Notice that the equality does not hold when vehicle $v_i$ transmits in any two subchannels $r_{u}$ and $r_{u'}$ that are in different subframes $l$ and $l'$, respectively. In a more general manner, for $N$ vehicles, this can be expressed as
\begin{equation} \label{e3}
	[({\bf I}_{N \times N} \otimes {\bf Q}_{L \times L}^{+}) {\bf x}_s] \circ [({\bf I}_{N \times N} \otimes {\bf Q}_{L \times L}^{-}) {\bf x}_s] = {\bf 0}_{NL \times 1}.
\end{equation}

The matrices ${\bf Q}^{+}$ and ${\bf Q}^{-}$ are boolean and contain significant information about the admissible and prohibited configurations regarding the time-dispersion of subchannels.

\subsection{Type IV: One-hop inter-cluster subchannel conflicts}
Two or more vehicles that are not aware of each other may transmit in the same subchannel. Thus, signals coming from these vehicles will merge and possibly become undecodable for other vehicles, specially for those that lie at the intersections of clusters. For any pair of vehicles $v_i \in \mathcal{V}^{(j)}$ and $v_{i'} \in \mathcal{V}^{(j')}$ located in different intersecting clusters but not at the intersection, the following must hold $x_{ik} x_{{i'}k} = 0 ~\forall r_k \in \mathcal{R}$ to prevent this kind of conflict. More generally, for $N$ vehicles these conditions can expressed as
\begin{equation} \label{e4}
[({\bf H}_{U \times N}^{+} \otimes {\bf I}_{KL \times KL}) {\bf x}] \circ [({\bf H}_{U \times N}^{-} \otimes {\bf I}_{KL \times KL}) {\bf x}] = {\bf 0}_{U \times 1}
\end{equation}
where $U$ is the number of vehicle pairs within one hop, e.g. $v_1$ and $v_9$ in Fig. \ref{f1}. The matrices ${\bf H}^{+}$ and ${\bf H}^{-}$ collect general knowledge of the vehicles that are within one-hop range that could potentially originate conflicts. 

The complete formulation of the problem is given by (\ref{e5}). Furthermore, to provide a better understanding of the matrices ${\bf G}^{-}$, ${\bf G}^{+}$, ${\bf Q}^{-}$, ${\bf Q}^{+}$, ${\bf H}^{-}$ and ${\bf H}^{+}$, consider the following.  
\begin{figure*}[!t]
	\begin{subequations} \label{e5}
		\begin{gather} 
		\small
		\begin{align}
		& {\rm max} ~ {\bf c}^T {\bf x} & \\
		& {\rm subject~to}~ \nonumber & \\ 
		& {\bf q}_{N \times 1} - {\bm \epsilon}\leq ({\bf I}_{N \times N} \otimes {\bf 1}_{1 \times KL})({\bf c}_{NKL \times 1} \circ {\bf x}_{NKL \times 1}) \leq {\bf q}_{N \times 1} + {\bm \epsilon} \\
		& [({\bf G}_{P \times N}^{+} \otimes {\bf I}_{L \times L}) ({\bf I}_{NL \times NL} \otimes {\bf 1}_{1 \times K}) {\bf x}] \circ [({\bf G}_{P \times N}^{-} \otimes {\bf I}_{L \times L})({\bf I}_{NL \times NL} \otimes {\bf 1}_{1 \times K}) {\bf x}] = {\bf 0}_{PL \times 1} \\
		& [({\bf I}_{N \times N} \otimes {\bf Q}_{L \times L}^{+}) ({\bf I}_{NL \times NL} \otimes {\bf 1}_{1 \times K}) {\bf x}] \circ [({\bf I}_{N \times N} \otimes {\bf Q}_{L \times L}^{-}) ({\bf I}_{NL \times NL} \otimes {\bf 1}_{1 \times K}) {\bf x}] = {\bf 0}_{NL \times 1} \\
		& [({\bf H}_{U \times N}^{+} \otimes {\bf I}_{KL \times KL}) {\bf x}] \circ [({\bf H}_{U \times N}^{-} \otimes {\bf I}_{KL \times KL}) {\bf x}] = {\bf 0}_{U \times 1}.
		\end{align}
		\end{gather}
	\end{subequations}
	\hrulefill
\end{figure*}

\vspace{0.25cm}
\textbf{Example:} Consider $N = 4$ vehicles distributed into $J = 2$ clusters, such that $\mathcal{V}^{(1)} = \{v_1, v_2, v_3\}$ and $\mathcal{V}^{(2)} = \{v_1, v_2, v_4\}$ with $\mathcal{V}^{(1)} \cap \mathcal{V}^{(2)} = \{v_1, v_2\}$. Also, $K = 3$ and $L = 3$. Thus, the matrices for this scenario are:
\tiny
\begin{gather} \nonumber
{\bf G}^{-} = 
\begin{bmatrix}
1 & 0 & 0 & 0 \\
1 & 0 & 0 & 0 \\
1 & 0 & 0 & 0 \\
0 & 1 & 0 & 0 \\
0 & 1 & 0 & 0 
\end{bmatrix}
{\bf G}^{+} = 
\begin{bmatrix}
0 & 1 & 0 & 0 \\
0 & 0 & 1 & 0 \\
0 & 0 & 0 & 1 \\
0 & 0 & 1 & 0 \\
0 & 0 & 0 & 1 \\
\end{bmatrix}
{\bf H}^{-} = 
\begin{bmatrix}
0 \\ 
0 \\
1 \\
0 
\end{bmatrix}
{\bf H}^{+} = 
\begin{bmatrix}
0 \\
0 \\
0 \\
1 
\end{bmatrix} \\
\nonumber
{\bf Q}^{-} = 
\begin{bmatrix}
0 & 0 & 0  \\
1 & 0 & 0 \\
0 & 1 & 1 
\end{bmatrix}
{\bf Q}^{+} = 
\begin{bmatrix}
1 & 0 & 0 \\
1 & 0 & 0 \\
0 & 1 & 0
\end{bmatrix}
{\bf Q}=[{\bf Q}^{-}]^T{\bf Q}^{+} = 
\begin{bmatrix}
0 & 0 & 0 \\
1 & 0 & 0 \\
1 & 1 & 0
\end{bmatrix}
\end{gather} 
\normalsize

The dimensions of ${\bf G}^{-}$ and ${\bf G}^{+}$ are $5 \times 4$ because there are $N = 4$ vehicles and $P = 5$ pairs of intra-cluster vehicles: $v_1-v_2$, $v_1-v_3$, $v_1-v_4$, $v_2-v_3$ and $v_2-v_4$. For instance, if $v_1$ and $v_2$ transmit concurrently, a Type II conflict will arise as they belong to the same cluster. This case is considered in row one of both matrices (first pair of vehicles), i.e., $[{\bf G}^{-}]_{11} = 1$ and $[{\bf G}^{+}]_{12} = 1$. However, $v_3$ and $v_4$ are in different clusters and the condition (\ref{e2}) is not violated. For this reason a row $p$ with such a combination where $[{\bf G}^{-}]_{p3} = 1$ and $[{\bf G}^{+}]_{p4} = 1$ does not exist. The square matrices ${\bf Q}^{-}$ and ${\bf Q}^{+}$ have dimensions $3 \times 3$ since there are $L=3$ subframes. The influence of these matrices can be best understood examining the product ${\bf Q}$. There are three combinations that lead to Type III conflicts, i.e. when any vehicle transmits in subframes 1 and 2 $([{\bf Q}]_{21} = 1)$, or 1 and 3 $([{\bf Q}]_{31} = 1)$, or 2 and 3 $([{\bf Q}]_{32} = 1)$ disregarding permutations. The square matrices ${\bf H}^{-}$ and ${\bf H}^{+}$ have dimensions $4 \times 1$ because there are $N = 4$  vehicles. Type IV conflicts will manifest when two vehicles that are at one hop distance transmit in the same subchannel. Such case happens when $v_3$ and $v_4$ transmit concurrently in time and frequency, i.e.  $[{\bf H}^{-}]_{31} = 1$ and $[{\bf H}^{+}]_{41} = 1$. 

\section{Proposed Subchannel Allocation Approach}
In this section, we realize the allocation of subchannels to vehicles following a three-stage process as depicted in Algorithm \ref{a1}. 

\begin{algorithm}[!b]
	\small
	\DontPrintSemicolon
	\Begin
	{
		\underline{Stage 1:}
		Sort the clusters in descending order of cardinality. 
		
		\For{$j = 1: J$}
		{
			\begin{tabular}{m{0.8cm} m{5.9cm}}
				\underline{Stage 2:} & Assign randomly to each vehicle $v_i \in \mathcal{V}^{(j)}$ \\ 
									 & some subframe $l_{k_i}$ without placing more than one vehicle in each subframe.\\
				
				\underline{Stage 3:} & Solve a knapsack problem for each vehicle \\ 
									 & $v_i \in \mathcal{V}^{(j)}$ \\ 
									 & \\ 
									 & \vspace{-0.2cm} \hspace{0.75cm} $\max \displaystyle \sum_{s = \{a \mid r_a \in \mathcal{R}_{k_i}\} } c_{is}$ \\ 
									 & \vspace{0.1cm} \hspace{0.75cm} subject to $\displaystyle \sum_{s = \{a \mid r_a \in \mathcal{R}_{k_i}\} } c_{is} \leq q_i$ \\
									 & \\ 
									 & \vspace{-0.2cm} where $\mathcal{R}_{k_i}$ is the set of subchannels in subframe $l_{k_i}$ .
			\end{tabular}
		}
	}
	\caption{Subchannel Allocation Algorithm based on Multiple Independent Knapsack Problems (MIKPs)}
	\label{a1}
\end{algorithm}

\underline{Stage 1:} The clusters are sorted hierarchically based on their cardinality such that $ \lvert \mathcal{V}^{(j)} \lvert \geq \lvert \mathcal{V}^{(j+1)} \lvert $. This means that clusters with a larger number of vehicles will be processed first. The intuitive reasoning is that larger clusters might be more difficult to optimize in terms of efficiently distributing the available subchannels among vehicles.

\underline{Stage 2:} Every vehicle in cluster $\mathcal{V}^{(j)}$ is randomly matched to some subframe without explicitly specifying the allocated subchannels. The only requirement is that each vehicle in $\mathcal{V}^{(j)}$ has to be assigned exactly one subframe to prevent Type III conflicts\footnote{A random pairing between vehicles and subframes is not the only possible manner of matching elements of these two groups. For instance, this could have been accomplished following a greedy or ordered criterion. Furthermore, an assignment based on the maximization of the achievable capacity per subframe could have been opted. Nevertheless, for the sake of simplification, in this work a random distribution was selected.}. In addition, no more than one vehicle in $\mathcal{V}^{(j)}$ can be placed in the same subframe to prevent Type II conflicts. Thus, a vehicle $v_i$ will broadcast over a set of subchannels located in some subframe $l_{k_i}$. 

\underline{Step 3:} Since vehicles have already been assigned to a specific subframe, a knapsack problem has to be solved to find which subchannels fulfill the QoS demands of the vehicle\footnote{The knapsack problem is solved through dynamic programming allowing to reduce the complexity compared to exhaustive search, which tests every possible combination of subchannels. In exchange of saving computation time, the knapsack problem requires a modest memory space to store previously computed partial combinations. The knapsack problem in this work is a special case known as the subset sum problem \cite{b3} \cite{b7}.}. The subchannel allocations that may cause vehicles to undergo Type IV conflicts are removed. Such information can be readily obtained from matrices ${\bf H}^{-}$ and ${\bf H}^{+}$. 

\section{Simulations}
\begin{figure}[!t]
	\centering
	\begin{tikzpicture}
	\begin{axis}[
	ybar,
	ymin = 0,
	ymax = 13.15,
	width = 9.2cm,
	height = 3.2cm,
	bar width = 20pt,
	tick align = inside,
	x label style={align=center, font=\footnotesize,},
	ylabel = {Rate [Mbps]},
	y label style={at={(-0.075,0.5)}, font=\footnotesize,},
	nodes near coords,
	every node near coord/.append style={font=\fontsize{7}{8}\selectfont},
	nodes near coords align = {vertical},
	symbolic x coords = {Average, Maximum, Minimum, Std. Dev.},
	x tick label style = {text width = 1.6cm, align = center, font = \footnotesize,},
	xtick = data,
	enlarge y limits = {value = 1.3, upper},
	enlarge x limits = 0.18,
	legend columns=-1,
	legend pos = north east,
	legend style={font=\fontsize{7}{6}\selectfont, text width=3cm,text height=0.02cm,text depth=.ex, fill = none, }]
	
	\addplot[fill = color0] coordinates {(Average,  12.18) (Maximum,13.15) (Minimum, 10.76) (Std. Dev., 0.72)}; \addlegendentry{Exact Formulation}
	
	\addplot[fill = color2] coordinates {(Average,  11.31) (Maximum, 13.15) (Minimum, 6.59) (Std. Dev., 1.13)}; \addlegendentry{MIKP-based Approach}

	\end{axis}
	\end{tikzpicture}
	\caption{Group of vehicles with target QoS = 12 Mbps and admissible range [10.4 - 13.6] Mbps}
	\label{f3}
	\vspace{-0.25cm}
\end{figure}
\begin{figure}[!t]
	\centering
	\begin{tikzpicture}
	\begin{axis}[
	ybar,
	ymin = 0,
	ymax = 10.17,
	width = 9.2cm,
	height = 3.2cm,
	bar width = 20pt,
	tick align = inside,
	x label style={align=center, font=\footnotesize,},
	ylabel = {Rate [Mbps]},
	y label style={at={(-0.075,0.5)}, font=\footnotesize,},
	nodes near coords,
	every node near coord/.append style={font=\fontsize{7}{8}\selectfont},
	nodes near coords align = {vertical},
	symbolic x coords = {Average, Maximum, Minimum, Std. Dev.},
	x tick label style = {text width = 1.6cm, align = center, font = \footnotesize,},
	xtick = data,
	enlarge y limits = {value = 1.3, upper},
	enlarge x limits = 0.18,
	legend columns=-1,
	legend pos = north east,
	legend style={font=\fontsize{7}{6}\selectfont, text width=3cm,text height=0.02cm,text depth=.ex, fill = none, }]
	
	\addplot[fill = color0] coordinates {(Average,  9.54) (Maximum, 10.17) (Minimum, 7.78) (Std. Dev., 0.59)}; \addlegendentry{Exact Formulation}
	
	\addplot[fill = color2] coordinates {(Average,  9.07) (Maximum, 10.17) (Minimum, 5.41) (Std. Dev., 1.02)}; \addlegendentry{MIKP-based Approach}
	
	\end{axis}
	\end{tikzpicture}
	\caption{Group of vehicles with target QoS = 9 Mbps and admissible range [7.4 - 10.6] Mbps}
	\label{f4}
	\vspace{-0.25cm}
\end{figure}
\begin{figure}[!t]
	\centering
	\begin{tikzpicture}
	\begin{axis}[
	ybar,
	ymin = 0,
	ymax = 7.18,
	width = 9.2cm,
	height = 3.2cm,
	bar width = 20pt,
	tick align = inside,
	x label style={align=center, font=\footnotesize,},
	ylabel = {Rate [Mbps]},
	y label style={at={(-0.075,0.5)}, font=\footnotesize,},
	nodes near coords,
	every node near coord/.append style={font=\fontsize{7}{8}\selectfont},
	nodes near coords align = {vertical},
	symbolic x coords = {Average, Maximum, Minimum, Std. Dev.},
	x tick label style = {text width = 1.6cm, align = center, font = \footnotesize,},
	xtick = data,
	enlarge y limits = {value = 1.3, upper},
	enlarge x limits = 0.18,
	legend columns=-1,
	legend pos = north east,
	legend style={font=\fontsize{7}{6}\selectfont, text width=3cm,text height=0.02cm,text depth=.ex, fill = none, }]
	
	\addplot[fill = color0] coordinates {(Average,  6.19) (Maximum, 7.18) (Minimum, 4.82) (Std. Dev., 0.71)}; \addlegendentry{Exact Formulation}
	
	\addplot[fill = color2] coordinates {(Average,  6.09) (Maximum, 7.18) (Minimum, 3.95) (Std. Dev., 0.82)}; \addlegendentry{MIKP-based Approach}
	
	\end{axis}
	\end{tikzpicture}
	\caption{Group of vehicles with target QoS = 6 Mbps and admissible range [4.4 - 7.6] Mbps}
	\label{f5}
	\vspace{-0.25cm}
\end{figure}
\begin{figure}[!t]
	\centering
	\begin{tikzpicture}
	\begin{axis}[
	ybar,
	ymin = 0,
	ymax = 5.99,
	width = 9.2cm,
	height = 3.2cm,
	bar width = 20pt,
	tick align = inside,
	x label style={align=center, font=\footnotesize,},
	ylabel = {Rate [Mbps]},
	y label style={at={(-0.075,0.5)}, font=\footnotesize,},
	nodes near coords,
	every node near coord/.append style={font=\fontsize{7}{8}\selectfont},
	nodes near coords align = {vertical},
	nodes near coords align = {vertical},
	symbolic x coords = {Average, Maximum, Minimum, Std. Dev.},
	x tick label style = {text width = 1.6cm, align = center, font = \footnotesize,},
	xtick = data,
	enlarge y limits = {value = 1.3, upper},
	enlarge x limits = 0.18,
	legend columns=-1,
	legend pos = north east,
	legend style={font=\fontsize{7}{6}\selectfont, text width=3cm,text height=0.02cm,text depth=.ex, fill = none, }]
	
	\addplot[fill = color0] coordinates {(Average,  3.79) (Maximum, 4.24) (Minimum, 2.08) (Std. Dev., 0.49)}; \addlegendentry{Exact Formulation}
	
	\addplot[fill = color2] coordinates {(Average,  3.89) (Maximum, 5.99) (Minimum, 1.23) (Std. Dev., 0.69)}; \addlegendentry{MIKP-based Approach}
	
	\end{axis}
	\end{tikzpicture}
	\caption{Group of vehicles with target QoS = 3 Mbps and admissible range [1.4 - 4.6] Mbps}
	\label{f6}
	\vspace{-0.25cm}
\end{figure}
By means of 1000 simulations, the performance of (\ref{e5}) and Algorithm \ref{a1} are evaluated using Matlab programming environment. We consider an scenario with $N = 40$ vehicles distributed over $J = 4$ clusters wherein $\lvert \mathcal{V}^{(1)} \lvert = 16$, $\lvert \mathcal{V}^{(2)} \lvert = 16$, $\lvert \mathcal{V}^{(3)} \lvert = 16$, $\lvert \mathcal{V}^{(4)} \lvert = 8$, and $\lvert \mathcal{V}^{(1)} \cap \mathcal{V}^{(2)} \cap \mathcal{V}^{(3)} \lvert = 8$, $\lvert \mathcal{V}^{(1)} \cap \mathcal{V}^{(4)} \lvert = \emptyset$, $\lvert \mathcal{V}^{(2)} \cap \mathcal{V}^{(4)} \lvert = \emptyset$, $\lvert \mathcal{V}^{(3)} \cap \mathcal{V}^{(4)} \lvert = \emptyset$. Furthermore, we assume that each vehicle requires any of the following QoS values $\{ 12, 9, 6, 3 \}$ Mbps and there are 10 vehicles of each kind spread across the clusters. The number of subframes is $L = 16$ and the number of subchannels per subframe is $K = 3$. Also, we consider $\epsilon = 1.6$ Mbps and the rate ranges are therefore $[ 10.4 - 13.6 ]$ Mbps, $[ 7.4 - 10.6 ]$ Mbps, $[ 4.4 - 7.6 ]$ Mbps and $[ 1.4 - 4.6 ]$ Mbps.

From Fig. \ref{f3} to Fig. \ref{f6} the achieved data rates for each group of vehicles are shown. Four criteria are employed to assess the performance of the two approaches. Although both approaches can in average provide the required QoS, it is critical to evaluate their performance in terms of the deviation from the target values. Thus, in the case of the MIKP-based approach, the target QoS is mostly not attained as the achieved rate values fall out of the admissible ranges. On the other hand, the exact formulation can guarantee the target QoS values in a tighter manner without severe deviation. Nevertheless, the disadvantage of the exact formulation is that a feasible solution may not be obtained if the requirements are not satisfied for all the vehicles. For the described setup, in 8\% of the cases a feasible solution was not found. This effect is not shown in the figures as only the successful allocations were considered. In practical situations, the parameter $\epsilon$ can be increased and thus the QoS requirements be relaxed for the algorithm to be executed again. However, this will depend on the particular scenario and whether the eNodeBs can afford to repeat the process. Although the proposed approach exhibits a simpler formulation and provides looser solutions, it can in 100\% of the cases provide at least a minimum operational level of service to all the vehicles. For this reason, under the criterion \textit{minimum}, the algorithm  does not perform as good as the exact formulation since every obtained solution contemplates that all in-coverage vehicles are always being served. Furthermore, the criterion \textit{standard deviation} depicts how different the attained QoS values are among vehicles with the same requirements. Thus, the proposed approach is less restrictive in this sense and the provided QoS might vary in a wider range. In both cases, no conflicts were originated although this outcome might change when the number of subchannels is scarcer.

\section{Concluding Remarks}
We presented a formulation for subchannel allocation in V2V \textit{mode-3} with four types of constrains. In addition, we proposed a simpler scheme using an extension of the knapsack problem in order to approach the initial formulation. Although the latter scheme is suboptimal, it does not fail in servicing and providing vehicles with subchannels because the QoS constraints are not as tight as in the original formulation.

\end{document}